\documentclass[aps,twocolumn,superscriptaddress,preprintnumbers]{revtex4}
\usepackage{epsfig}
\usepackage{graphicx,amssymb,epsfig,amsmath}
\usepackage{color}
\usepackage[normalem]{ulem}
\def\<{\langle}
\def\>{\rangle}
\def\ben{\begin{eqnarray}}
\def\een{\end{eqnarray}}

\def\bei{\begin{itemize}}
\def\eei{\end{itemize}}

\def\Bell{{\rm Bell}}

\newcommand{\be}{\begin{eqnarray}}
\newcommand{\ee}{\end{eqnarray}}

\newcommand{\braopket}[3]{\langle #1|#2|#3 \rangle}

\newcommand{\removed}[1]{}
\renewcommand{\eqref}[1]{(\ref{#1})}

\begin{document}
\title{Bell measurement rules out supraquantum correlations}

\author{L. Czekaj}
\email{jasiek.gda@gmail.com}
\affiliation{Faculty of Applied Physics and Mathematics, National Quantum Information Centre, Gda\'nsk University of Technology, 
80-233 Gda\'nsk, Poland}

\author{M. Horodecki}
\affiliation{Institute of Theoretical Physics and Astrophysics, National Quantum Information Centre, Faculty of Mathematics, Physics and Informatics, University of Gda{\'n}sk, Wita Stwosza 57, 80-308 Gda{\'n}sk, Poland}

\author{T. Tylec}
\affiliation{Institute of Theoretical Physics and Astrophysics, University of Gda\'nsk, 80-952 Gda\'nsk, Poland}

\date{\today}
\begin{abstract}

The so called bipartite non-signaling boxes are systems whose statistics is constrained solely by
the principle of no instantaneous signaling between distant locations. Such systems can exhibit  much stronger correlations than those 
admitted by quantum mechanics. Inspired by quantum logic approach  of Tylec and Ku\'{s},  J. Phys. A: Math. Theor. 48 (2015) 505303, we consider non-signaling boxes with three inputs per party, and extend the set of measurements with just a single {\it global} measurement - one that mimics quantum two-party Bell measurement. We then show that this seemingly mild extension completely rules out supraquantum correlations: the resulting system admits precisely quantum mechanical correlations of two qubits. We also consider non-maximally entangled measurements, obtaining  interpolation between quantum and full no-signaling theory. Our study paves a way  to a general programme of
amending no-signaling theories with some measurements inherited from quantum mechanics, leading to various  interpolations between 
non-signaling boxes and quantum mechanics. 
\end{abstract}

\maketitle
\section{Introduction}
The idea of non-signaling boxes introduced in \cite{PR} has become very fruitful for several purposes. 
First, the non-signaling boxes have found an application to device-independent cryptography \cite{Barrett_crypto_Bell}, basing solely on assumption of no instantaneous signaling and violation of Bell inequalities \cite{Bellreview}. In particular, randomness amplification and expansion 
protocol have been proposed \cite{Pironio_rand_exp_Nature,ColbeckRenner}, whose verification does not require any knowledge on quantum mechanics, 
but can be done based  solely on statistical behaviour of the devices. On the other hand, analysis of non-signaling boxes leads to a better understanding of capabilities and limitations of quantum theory itself: the set of quantum mechanical states forms a convex body situated between two polytopes: the classical polytope of Kolmogorovian probability distributions and the larger poytope of all non-signaling boxes. 
The concept of non-signaling boxes has lead to a vast field of so called General Probabilistic Theories (GPT) \cite{GPT1, GPT2, GPT3}. 


Recently, the relations between GPT and quantum logic were analysed~\cite{TylecKus}.
The authors use the framework of quantum logic to construct logic of propositions of two-party non-signaling boxes. They build the logic from propositions describing single party and prove that the logic indeed describe spatially separated subsystems.



So far within the subject of GPT, not much has been done regarding joint measurements on composite systems. 
Boxes with bipartite measurements were considered in~\cite{hypersigVedral} and were used to construct examples of theories violating "no-hypersignaling principle" formulated therein. These measurements were based on extremal points of the no-signaling polytope. The resulting models either exhibit solely classical correlations (when measurements corresponding to all extremal points are added) or exhibit maximally non-local correlations 
- those violating Tsirelson bound \cite{Tsi}. 
In this context, an important challenge is to build  models that interpolate  between  the two extremes. 
For single system, an important example of such interpolation is family of polygonic models \cite{Massar-polygon} some  of them violating Holevo bound. A bipartite models based on polygonic local systems was also considered \cite{polygon-CHSH}, some of them violating Tsirelson bound. Yet, the joint measurements possible for those systems have not been analyzed. 


In this paper we want to  to avoid  the binary situation: classical or full nosingaling, hence we need more sophisticated measurements than ones used in \cite{hypersigVedral}. To this end  we take inspiration from the quantum logic approach to nosignaling boxes of Ref.  \cite{TylecKus}. We aim to analyse the effect of enriching the initial model - a standard no signaling-box, which admits just product measurements - 
with an {\it entangled measurement inherited from quantum mechanics}.  

The basic global measurement in quantum mechanics one may think of is clearly the Bell measurement  \cite{MBRevzen_Bell}. 
Surprisingly,  we obtain that adding just this single measurement severely constrains the set of possible states. 
Namely, we show that the {\it no-signaling box with Bell measurement exhibits no supraquantum correlations}. It actually reproduces exactly 
all quantum correlations. We do it by showing that existence of Bell measurement, combined with natural assumption, that product of local states is a legitimate joint state, imposes that local states form a ball, i.e. it is the same as the set of states of qubit. 
Then we use the result of \cite{LQandNS}
where it is shown, that bipartite systems which are locally quantum, and are no-signaling, admit only quantum correlations. 

We also consider non-maximally entangled measurements, and obtain interpolation between local systems being balls (like in quantum mechanics) and cubes (i.e completely unrestricted local systems).



\section{Model}
We consider system $\mathcal{S}_{AB}$ composed of two elementary subsystems $\mathcal{S}_A, \mathcal{S}_B$. State spaces of elementary systems $\mathcal{S}_A$ and $\mathcal{S}_B$ are identical. 

The elementary system may be measured by means of one of three dichotomic measurements $X, Y, Z$. The measurements are not compatible, i.e. they cannot be measured together. In this sense they mimic Pauli measurements for quantum system. However at this step we do not put any constraints on the measurements outcomes probabilities beside standard positivity and normalisation constraints. 
In particular, there are no uncertainty constraints for elementary system.

The state of the elementary system $\mathcal{S}_A$ is described by probabilities of measurement outcomes:
$p(a|x_A)$, where $x_A\in\{X,Y,Z\}$ enumerates measurements and $a$ enumerates outcomes "$+$", "$-$" (similarly for $\mathcal{S}_B$).

Now we move to composed system $\mathcal{S}_{AB}$. 
Consider first standard  no-signaling bipartite boxes \cite{PR}. These boxes are described by probabilities $p(ab|x_A x_B)$ where $a\in\{+,-\}$ denote output of measurement $x_A\in\{X,Y,Z\}$ performed on subsystem $A$, analogically for $b$ and $x_B$.
Probabilities $p(ab|x_A x_B)$ fulfill non-signaling conditions, i.e. probability of outcomes of measurement performed on subsystem $A$ do not depend on the measurement performed on the subsystem $B$ (analogically for $B$ and $A$). The condition is expressed by equation:
\begin{equation}
\sum_{b} p(a,b|x_A,x_B) = \sum_{b} p(a,b|x_A,x_B'),
\end{equation}
which holds for all $a\in\{+,-\}$ and $x_A,x_B,x_B'\in\{X,Y,Z\}$. 

So far this is a standard "no-signaling box". We shall now assume, that there is additional 2-party measurement which cannot be represented as a joint measurement of two local measurements. This intrinsically 2-party measurement returning one of the 4 outcomes $k=1,2,3,4$. 
We will define probabilities of these outcomes using parity relations for Bell measurement known from quantum mechanics,
hence the probabilities will be denoted by $p(k|\Bell)$, and the measurement we will call "Bell measurement". 

In quantum mechanics we we have 
\begin{eqnarray}
|\phi^+\>\<\phi^+| +|\psi^+\>\<\psi^+| &=&P_A^{X,+}\otimes P_B^{X,+}+P_A^{X,-}\otimes P_B^{X,-}\nonumber \\
|\phi^-\>\<\phi^-| +|\psi^+\>\<\psi^+| &=&P_A^{Y,+}\otimes P_B^{Y,+}+P_A^{Y,-}\otimes P_B^{Y,-}\nonumber \\
|\phi^+\>\<\phi^+| +|\phi^-\>\<\phi^-| &=&P_A^{Z,+}\otimes P_B^{Z,+}+P_A^{Z,-}\otimes P_B^{Z,-}\nonumber \\
\end{eqnarray}
where 
\ben
&&\phi_\pm = \frac{1}{\sqrt 2} (|0\>|0\>\pm|1\>|1\>) \nonumber \\
&&\psi_\pm = \frac{1}{\sqrt 2} (|0\>|1\>\pm|1\>|0\>)
\een
and $P_A^{X,\pm}$,  $P_A^{Y,\pm}$,$P_A^{Z,\pm}$ are eigenprojectors of Pauli matrices $\sigma_x,\sigma_y, \sigma_z$ respectively
(same for $B$).  
We now impose the same relations for our joint measurement, on the level of statistics: 
\begin{eqnarray}
&&p(1|\Bell)+p(3|\Bell)=p(++|XX)+p(--|XX)\nonumber \\
&&p(2|\Bell)+p(3|\Bell)=p(++|YY)+p(--|YY)\nonumber \\
&&p(1|\Bell)+p(2|\Bell)=p(++|ZZ)+p(--|ZZ).\nonumber\\
\label{eq:bellProbsDef}
\end{eqnarray}
Notice, that it would not make sense to impose non-signaling conditions onto "Bell measurement" because the latter is performed on the whole
system. The system will be now fully described by the set of probabilities $p(ab|x_A x_B)$ and $p(k|\Bell)$.

It is worth to mention that state space of the composed system without "Bell measurement" is maximal tensor product space~\cite{maxTensor} 
and with the elementary system as described above, state space of such composed system is full non-signaling polytope \cite{Barrett_PR_boxes}. 
In the further part of the paper we will show that equipping composed system with "Bell measurement" will change this picture a lot. 

State representation with probabilities $p(ab|x_Ax_B)$ and $p(k|\Bell)$ contains 40 parameters. However they are not independent. Using non-signaling conditions together with normalization and relation~\eqref{eq:bellProbsDef} we can express state of the composed system 
using $15$ free parameters: probabilities of positive outcomes for every measurement settings $p(++|x_Ax_B)$ and marginal probabilities $p(+|x_A),p(+|y_B)$ (due to non-signaling condition, we can write marginal probability as $p(+|x_A) = p(++|x_AX)+p(+-|x_AX)$). We can arrange these parameters in form of matrix:
\begin{equation}
\left(\begin{array}{cccc}
p(++|XX)&\cdots&\cdots&p(+|X_B)\\
\vdots&p(++|YY)&\cdots&p(+|Y_B)\\
\vdots&\vdots&p(++|ZZ)&p(+|Z_B)\\
p(+|X_A)&p(+|Y_A)&p(+|Z_A)&1
\end{array}\right).
\label{eq:stateRepresentationShort}
\end{equation}
In particular, the state will is fully determined by the statistics of local measurements satisfying therefore 
local tomography principle \cite{localTomographyB, localTomographyC}. 

Before we move further in analysis of state space $\Omega_{AB}$, let us make a digression. Namely, suppose that the elementary system is equipped only with two  measurements $X,Y$. One then finds, that in such theory, extending the set of measurements 
with the "Bell measurement" leads to additional free parameter. It follows from fact that we can write only the first two equations 
from~\eqref{eq:bellProbsDef}. This theory does not fulfill
local tomography principle \cite{localTomographyB, localTomographyC}, since state of composed system cannot be fully described in terms of joint probabilities of local measurements, i.e. in terms of $p(ab|x_A,x_B)$. This is analogous to the difference between complex and real quantum mechanics where local tomography is a crucial piece \cite{localTomographyA}. 

\section{Constraints for correlations imposed by existence of "Bell measurement"}
In this section we will show that correlations exhibited by boxes admitting "Bell measurement" are exactly the quantum ones.
To this end we will study how conditions imposed by existence "Bell measurement" impacts state space of elementary system.
We shall assume that two natural conditions hold: 
\bei
\item[(i)] the sets of states of local systems are the same, i.e.   $\Omega_A=\Omega_B$
\item[(ii)] all product states are allowed, i.e.  $\Omega_A\otimes\Omega_B\subset\Omega_{AB}$.
\eei
We shall now show that non-negativity of $p(k|\Bell)$ together with these assumptions  leads to equivalence of elementary system state space with Bloch ball. Then, knowing that elementary state space is quantum and the composed system is non-signaling, we can directly use results from~\cite{LQandNS} to obtain that all correlations in bipartite system are quantum correlations.


To proceed, consider product of two identical states  $\omega_{AB}=\omega_A\otimes\omega_B$ which 
by assumption (ii) is allowed. 
Denote marginals by  
\ben
&&p(+|x_A=X)=p(+|x_B=X)=p_X, \nonumber \\
&&p(+|x_A=Y)=p(+|x_B=Y)=p_Y, \nonumber \\
&&p(+|x_A=Z)=p(+|x_B=Z)=p_Z
\ee 
and consider probability $p(4|\Bell)$ for that state. Then, from simple algebra, we get that:
\begin{eqnarray}
&&p(4|\Bell)= \nonumber \\
&&=\frac{1}{2}\bigl(p(+|x_A=X)+p(+|x_A=Y)+p_A(+|x_A=Z)\nonumber \\
&&+\,p(+|x_B=X)+p(+|x_B=Y)+p_B(+|x_B=Z)\nonumber  \\
&&-\,p(++|XX)-p(++|YY)-p(++|ZZ)-1\bigr)
\end{eqnarray}
and for state $\omega_{AB}$:
\begin{eqnarray}
p(4|\Bell)&=&p_X+p_Y+p_Z-p_X^2-p_Y^2-p_Z^2-\frac{1}{2}.
\end{eqnarray}
We can rewrite the above expression together with positivity condition for $p(4|\Bell)$ as:
\begin{equation}
\left(p_X-\frac{1}{2}\right)^2+\left(p_Y-\frac{1}{2}\right)^2+\left(p_Z-\frac{1}{2}\right)^2\leq\left(\frac{1}{2}\right)^2.
\label{eq:ballDef}
\end{equation}
This formula constrains state space of elementary system and  we see, that the condition is equivalent to Bloch ball  for averages of 
observables $X,Y,X$. In particular, the constraints can be interpreted as uncertainty relation expressed in terms of probability of measurement outcome.  

Now, we will show, that these constraints are tight, i.e. that all the product states products of states fulfilling~\eqref{eq:ballDef} give positive values for the outcomes probabilities of measurement. Of course, for products of local measurements, 
the product states give positive probabilities by definition. So we need to check whether they give positive values 
of probabilities of outcomes just for  the "Bell measurement". 

To this end we rewrite positivity conditions for $p(k|\Bell)$ in moments representation (i.e. mean values of local measurement, e.g. $m_X^A = \frac{1}{2}\left(p(+|x_A=X) - p(-|x_A=X)\right)=p(+|x_A=X)-\frac{1}{2}$). 
When  we arrange moments in form of vector $m=(m_X,m_Y,m_Z,1)$, then positivity condition take form: 
\begin{equation}
0\leq\braopket{m^A}{T_k}{m^B}\label{posCondOp},
\end{equation}
where $T_k$ are diagonal  matrices representing outcomes $k$ given by 
\begin{eqnarray}
T_1&=&\frac{1}{4}\text{diag}(1,-1,1,1),\nonumber \\
T_2&=&\frac{1}{4}\text{diag}(-1,1,1,1),\nonumber \\
T_3&=&\frac{1}{4}\text{diag}(1,1,-1,1),\nonumber \\
T_4&=&\frac{1}{4}\text{diag}(-1,-1,-1,1),
\end{eqnarray}
where $\text{diag}(\ldots)$ denotes diagonal matrix with given entries.

Formula~\eqref{posCondOp} can be unwind to:
\begin{eqnarray}
k=1&:& -m_X^A\cdot m_X^B + m_Y^A\cdot m_Y^B - m_Z^A\cdot m_Z^B \leq r^2,\label{eq:bellAsMoment}\\
k=2&:& -m_X^A\cdot m_X^B - m_Y^A\cdot m_Y^B + m_Z^A\cdot m_Z^B \leq r^2,\\
k=3&:&m_X^A\cdot m_X^B - m_Y^A\cdot m_Y^B - m_Z^A\cdot m_Z^B \leq r^2,\\
k=4&:&m_X^A\cdot m_X^B + m_Y^A\cdot m_Y^B + m_Z^A\cdot m_Z^B \leq r^2,\label{eq:psiDotProd}
\end{eqnarray}
where $r=\frac{1}{2}$. 
First observe that LHS of~\eqref{eq:psiDotProd} has a form of scalar product between vectors $(m^A_X,m^A_Y,m^A_Z)$ and $(m^B_X,m^B_Y,m^B_Z)$. We know from~\eqref{eq:ballDef} that norm of these vectors is bounded by $r$, therefore~\eqref{eq:psiDotProd} holds for all states from ball given by~\eqref{eq:ballDef}. The other inequalities can be easily translate to the form of scalar product: because of symmetry of state space we can always replace state on $\mathcal{S}_B$ by the state with appropriate observable flipped.   
We can conclude that all states given by~\eqref{eq:ballDef} fulfill positivity constraints, therefore~\eqref{eq:ballDef} define state space of local system and in fact it is Bloch ball. 

As said, knowing that elementary state space is quantum and composed system is non-signaling, we can directly use results from~\cite{LQandNS} to say that all correlations in bipartite system are quantum correlations.

\section{Adding new measurements: quantum logic approach versus local tomographic approach. }
The way we approached the definition of Bell measurement for a non-signaling box 
was to  enforce the relation between statistics of the new measurement and the statistics of the local measurements to be the same as the relation between statistics of Pauli measurements and Bell measurement in quantum mechanics. The inspiration was taken from quantum logic approach to non-signaling boxes. 

{\it The "Bell measurement" and quantum logic approach}
One starts from the logic structure of non-signaling boxes. The paper~\cite{TylecKus} provides the set of valid propositions for non-signaling boxes. Example of valid the proposition is "the system is in state $++$ of measurement $XX$". More over we know that the proposition "the system is in state $++$ or $--$ of measurement $XX$" is also valid. These two propositions refers to probabilities $p(++|XX)$ and $p(++|XX)+p(--|XX)$ for given boxes. In contrast, the proposition "the system is in state $++$ of measurement $XX$" or $++$ of measurement $ZZ$" is not valid. This works in analogy to algebra of orthogonal projectors in quantum mechanics. 
Now one observes that some propositions in quantum mechanic may be expressed in several ways: e.g. parity $XX$ may be expressed as $P_A^{X,+}\otimes P_B^{X,+}+P_A^{X,-}\otimes P_B^{X,-}$ or $|\phi^+\>\<\phi^+| +|\psi^+\>\<\psi^+|$.
We require the same type of relations to hold in our model. That leads to equation~\eqref{eq:bellProbsDef}. 

{\it Adding measurements via local tomographic approach.}
The considered definition of "Bell measurement"  can be seen as an instance of a more general way of inheriting joint measurements from quantum theory  that is not covered by quantum logic approach - the one based on local tomography. Namely, suppose that we consider some measurement from quantum mechanics and want to impose it onto a no-signaling box. In quantum mechanics,  
due to local tomography, the statistics of the local observables determine statistics of all measurements. 
We can thus define a new measurement by requiring that statistics of its outcomes to be determined by the statistics of local observables 
through the quantum mechanical relation. 
It is then possible to extend no-signaling boxes with analogue of quantum measurement in non maximally entangled basis.

\section{Noisy Bell measurement and non-maximally entangled measurement}
\begin{figure}[h]
	\centering
		\includegraphics[scale=0.4]{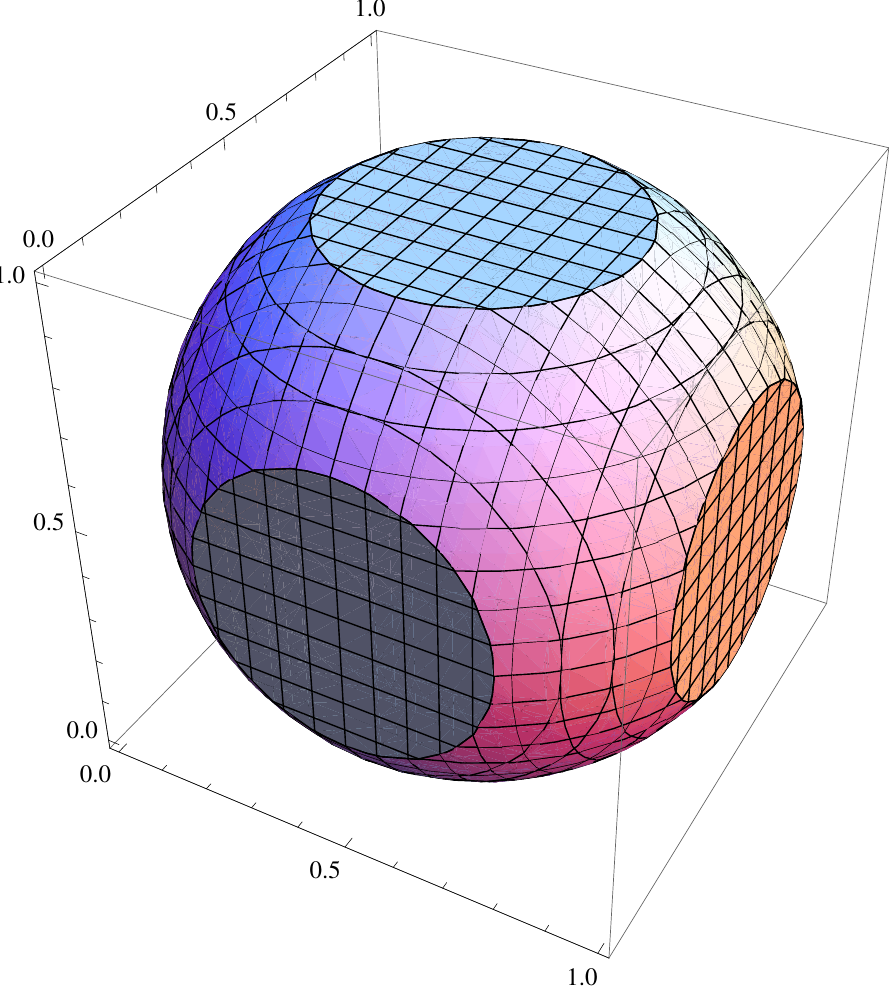}
		\includegraphics[scale=0.4]{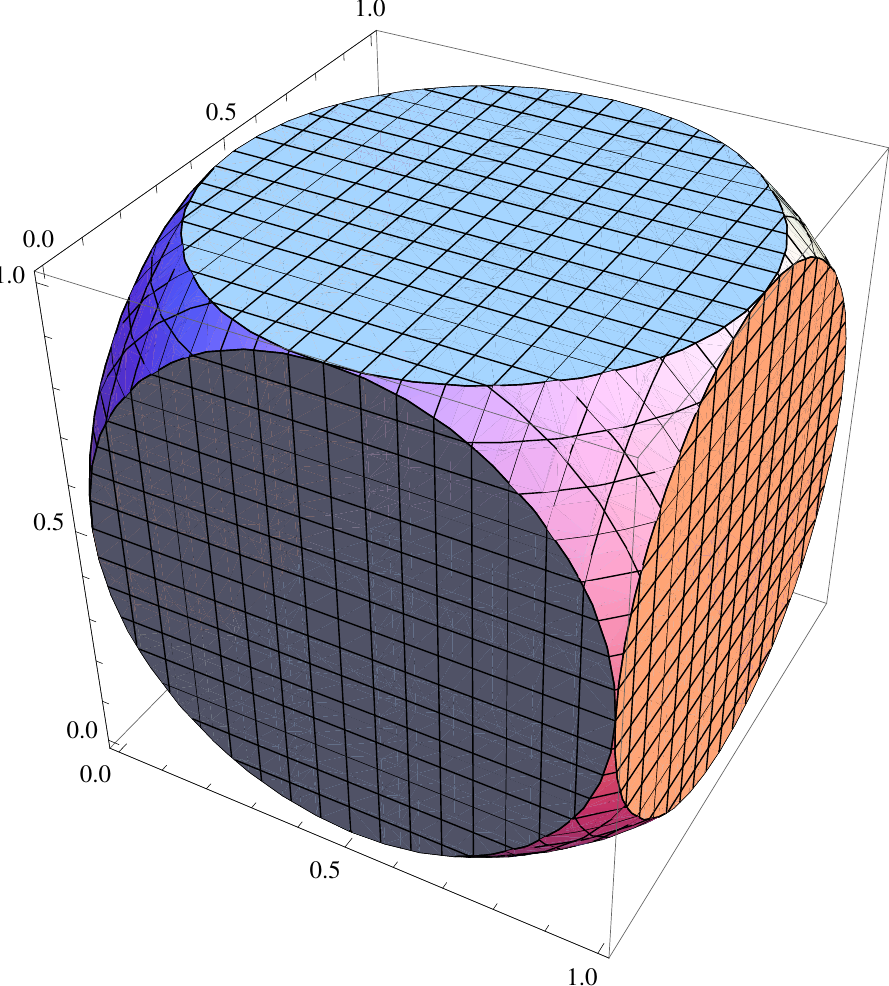}
	\caption{\label{fig:stateSpaceNoise}
	State space for noisy Bell measurement for $\lambda=1/4$ (left) and $\lambda=1/2$ (right).}
\end{figure}

Here we will present how noisy Bell measurement as well as  measurement in a non-maximally entangled basis modify 
the locla state space. We shall use local tomography approach to define these measurements for no-signaling boxes.
{\it Noisy Bell measurement.}
We will consider here measurement inherited from POVM with elements 
\ben
&&(1-\lambda)|\phi_\pm\>\<\phi_\pm| + \lambda I/4,\nonumber\\ 
&&(1-\lambda)|\psi_\pm\>\<\psi_\pm| + \lambda I/4, 
\een
The probabilities $p(k|\widetilde{Bell})$ of the measurements are related to the probabilities $p(k|Bell)$ as follows:
\begin{equation}
p( k| \widetilde{Bell}) = (1-\lambda)p(k|Bell) + \lambda/4,
\end{equation}
and can be expressed in the terms of moments by~\eqref{eq:bellAsMoment}-\eqref{eq:psiDotProd} with $r=\frac{1}{2\sqrt{1-\lambda}}$.
The argumentation analogical to the one present in case of "Bell measurement" can be used here to show that state space of elementary system (in terms of probabilities) is $1/2$ centered ball with $r=\frac{1}{2\sqrt{1-\lambda}}$ restricted to the box of $0\leq p_X,p_Y,p_Z \leq 1$ (see FIG.~\ref{fig:stateSpaceNoise}).

{\it Non-maximally entangled measurement.}
Here we take non-maximally entangled basis parametrized by $a$ and $b$ 
\ben
&&\tilde\phi_\pm = \frac{1}{\sqrt 2} (a|0\>|0\>\pm b|1\>|1\>) \nonumber \\
&&\tilde\psi_\pm = \frac{1}{\sqrt 2} (a|0\>|1\>\pm b|1\>|0\>).
\een
Taking $a=0, b=1$ leads to product basis and $a=b=1/\sqrt{2}$ to standard Bell basis. 

We express positivity conditions in terms of formula~\eqref{posCondOp} (outcome $0$ refers to $\tilde{\phi}_+$, $1$ to $\tilde{\phi}_-$, etc.):
\begin{eqnarray}
T_1 &=& \frac{1}{4}
\left(\begin{array}{cccc}
2ab&0&0&0\\
0&-2ab&0&0\\
0&0&1&a^2-b^2\\
0&0&a^2-b^2&1
\end{array}\right),\\
T_2 &=& \frac{1}{4}
\left(\begin{array}{cccc}
-2ab&0&0&0\\
0&2ab&0&0\\
0&0&1&a^2-b^2\\
0&0&a^2-b^2&1
\end{array}\right),\\
T_3 &=& \frac{1}{4}
\left(\begin{array}{cccc}
2ab&0&0&0\\
0&2ab&0&0\\
0&0&-1&a^2-b^2\\
0&0&-a^2+b^2&1
\end{array}\right),\\
T_4 &=& \frac{1}{4}
\left(\begin{array}{cccc}
-2ab&0&0&0\\
0&-2ab&0&0\\
0&0&-1&a^2-b^2\\
0&0&-a^2+b^2&1
\end{array}\right).
\end{eqnarray}

In the following part we will analyze state space for particular basis with $a=\sin(\pi/4 + \alpha)$, $b=\cos(\pi/4 + \alpha)$. For that parametrization we can write~\eqref{posCondOp} as (cf.~\eqref{eq:bellAsMoment}-\eqref{eq:psiDotProd}):
\begin{eqnarray}
- m^A_Z m^B_Z + ( -m^A_X m^B_X + m^A_Y m^B_Y) \cos(2\alpha) &&\nonumber\\
- (m^A_Z + m^B_Z) \sin(2 \alpha) &\leq& 1,\nonumber\\
- m^A_Z m^B_Z + ( m^A_X m^B_X - m^A_Y m^B_Y) \cos(2\alpha) &&\nonumber\\
+ (m^A_Z + m^B_Z) \sin(2 \alpha) &\leq& 1,\nonumber\\
m^A_Z m^B_Z - ( m^A_X m^B_X + m^A_Y m^B_Y) \cos(2\alpha) &&\nonumber\\
- (m^A_Z - m^B_Z) \sin(2 \alpha) &\leq& 1,\nonumber\\
m^A_Z m^B_Z + ( m^A_X m^B_X + m^A_Y m^B_Y) \cos(2\alpha) &&\nonumber\\
+ (m^A_Z - m^B_Z) \sin(2 \alpha) &\leq& 1.\nonumber
\end{eqnarray}

\begin{figure}[h]
	\centering
		\includegraphics[scale=0.45]{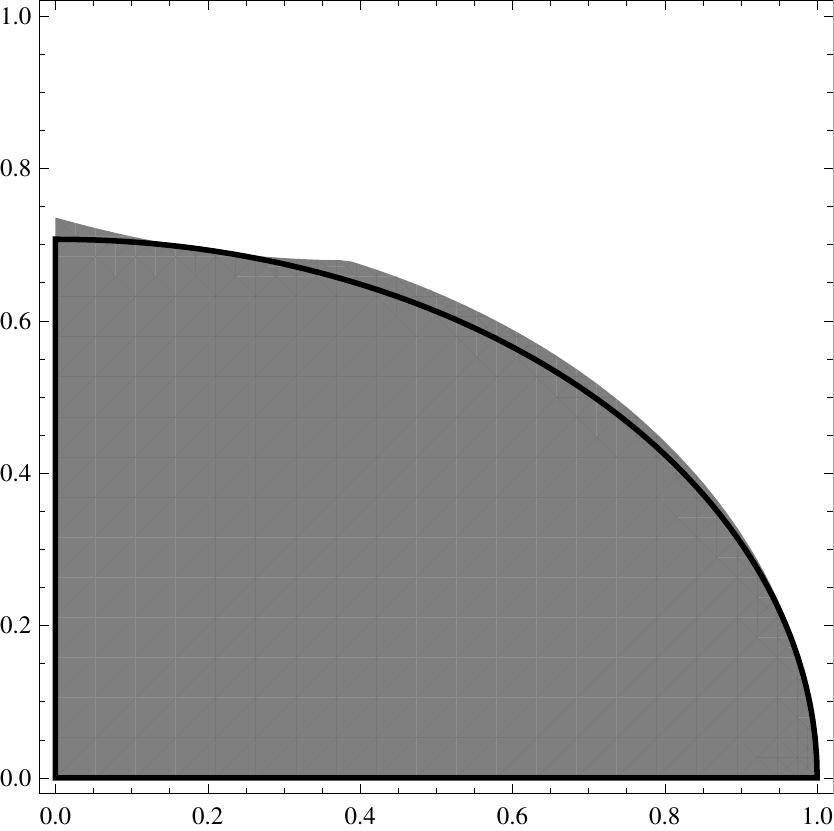}
		\includegraphics[scale=0.45]{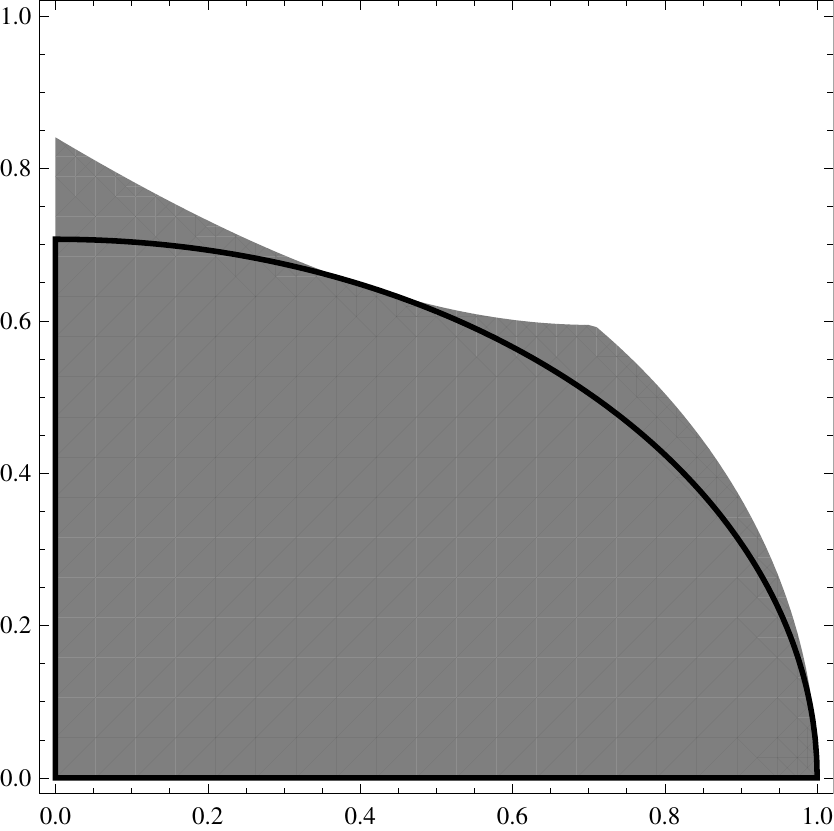}
		\includegraphics[scale=0.45]{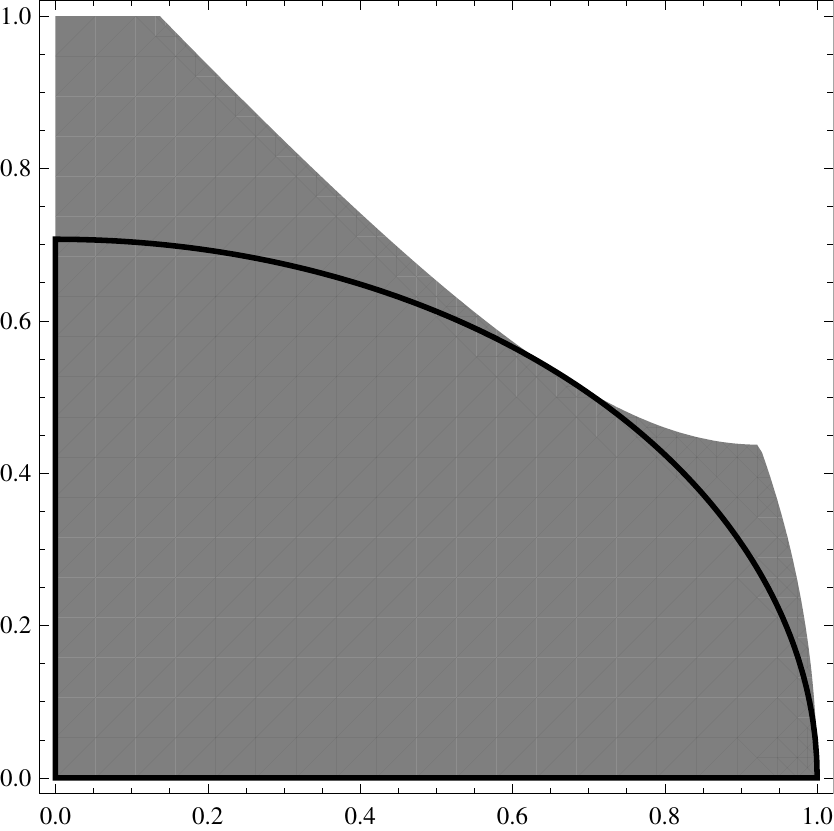}
		\includegraphics[scale=0.45]{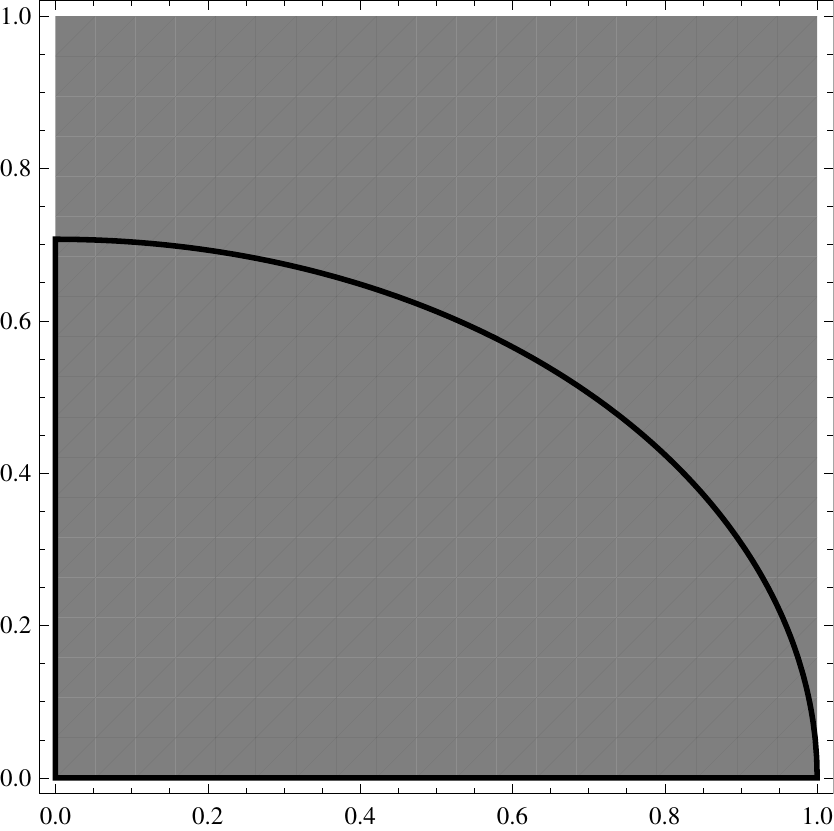}
	\caption{\label{fig:nonMaxEnt}
	Allowed values of $l$ (OY) and $h$ (OX) for $\alpha=\pi/16, 2\pi/16, 3\pi/16, 4\pi/16$. Thick black line bounds region permitted in quantum mechanics. We can observe interpolation between quantum and unrestricted system.}
\end{figure}

It is hard to obtain full state space for given positivity conditions. Moreover there may be many inequivalent states spaces. Here we are interested in interpolation between quantum and unrestricted system. For this reason we bound state space of single system from inside by cube in moments representation with vertices $(\pm l,\pm l, \pm h)$. Because of linearity it is enough to check if vertices fulfill positivity conditions. For Bell basis ($\alpha = 0$ which leads to $a=b=1/\sqrt{2}$) we get condition:
\begin{equation}
2 l^2 + h^2 \leq 1.
\end{equation}
On FIG.~\ref{fig:nonMaxEnt} we present permitted values of $l$ and $h$ for different parameter $\alpha$.

\section{Concluding remarks}
Our model of no-signaling boxes admitting Bell measurement or non-maximally entangled measurement is just an example of 
constraining the nosignaling theory by amending it by quantum-inherited joint measurements. 
The results encourage to study other amendments, and checking their properties. In particular, 
it is worth to examine multipartite systems where quantumness of local systems does not determine the correlations anymore 
\cite{tripartite_local_quantum}. 
Another route is to consider more general parity measurements, e.g with more outcomes than just four, in place of Bell measurements. 
Finally it would be interesting to perform more detailed study of the correlations exhibited by no-signaling systems with non-maximally 
entangled measurements. Such analysis involves a highly nonlinear problem, which requires further investigations. 


 {\it Acknowledgements.}
\L{}Cz and MH are supported by John Templeton Foundation.



\end{document}